*Article*

# CALPHAD-based modelling of the temperature-composition-structure relationship during physical vapor deposition of Mg-Ca thin films

Philipp Keuter [1,2,*], Moritz to Baben [1], Shamsa Aliramaji[2] and Jochen M. Schneider [2]

[1] GTT-Technologies, Kaiserstraße 103, Herzogenrath 52134, Germany; mtb@gtt-technologies.de
[2] Materials Chemistry, RWTH Aachen University, Kopernikusstr. 10, 52074 Aachen, Germany; aliramaji@mch.rwth-aachen.de, schneider@mch.rwth-aachen.de
* Correspondence: pk@gtt-technologies.de

**Abstract:** The temperature-dependent composition and phase formation during physical vapor deposition (PVD) of Mg-Ca thin films is modelled using a CALPHAD-based approach. Considering the Mg and Ca sublimation fluxes calculated based on the vapor pressure obtained by employing equilibrium thermochemical calculations, experimentally observed synthesis temperature trends in thin film composition and phase formation are reproduced. The model is a significant step towards understanding how synthesis parameters control composition and thereby phase formation in PVD of metals with high vapor pressures.

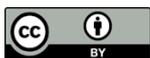

## 1. Introduction

Physical vapor deposition (PVD) processes, e.g. magnetron sputtering, usually lead to non-equilibrium phase formation in the as-grown thin films[1]. In recent years, however, it has been demonstrated that thin film phase formation is closer to (at least: para-) equilibrium than often assumed. Spencer has demonstrated that robust thermodynamic descriptions of metastable phases can be used to predict phase formation both in metallic and nitride systems already more than 20 years ago[2]. More recently, Chang et al. have extended the underlying model assumptions by a numerical treatment of the surface diffusion kinetics relative to the deposition rate[3–5]. For reactive deposition, to Baben et al. have shown that a para-equilibrium assumption, i.e. assuming equal chemical potential of nitrogen in the gas phase and in the deposited thin film, can be used to understand off-stoichiometry of as deposited $(Ti,Al)N_x$ thin films[6]. Here, we extend this further and consider the kinetics of metal sublimation at the example of magnetron sputtered Mg-Ca thin films.

Mg sublimation during sputtering has, for example, been observed for co-sputtering of Al-Mg-B thin films at substrate temperatures of 600 °C[7] limiting the temperature range for synthesis. The same effect has been observed also for other metals exhibiting high vapor pressure at deposition temperature such as Al[8], Mn[9], Sn[10], and Ge[10,11]. A pronounced variation in Mg-concentration depending on the synthesis temperature has also been reported for Mg-Ca thin films where minor variations in synthesis temperature yield pronounced modifications in thin film composition[12], thus, revealing a complex synthesis-composition-structure relationship when Mg is involved. Even though PVD is frequently applied to study the properties of elemental Mg[13,14] and Mg-based films[7,12,15–22], the sublimation behavior of Mg has only been reasoned qualitatively based on ground-state density function theory calculations[12]. In this work, equilibrium thermochemical calculations employing the CALPHAD approach are used to develop a predictive model of the temperature-dependent phase formation during Mg-Ca thin film growth. In this model, sublimation is considered as the dominating desorption mechanism. Remaining deviations obtained between experimental results and calculations within the temperature range for phase pure Mg$_2$Ca formation are further investigated by correlative structural and chemical analyses of annealed combinatorial Mg-Ca thin films supporting a wider stoichiometry range of Mg$_2$Ca than expected from the CALPHAD model.

## 2. Materials and Methods

### 2.1 Theoretical methods

Equilibrium thermochemical calculations were performed using the Equilib module of FactSage[23] version 8.1 with the commercial FTlite[24] database for the solid and liquid phases and FactPS[23] for the ideal gas phase. During magnetron sputtering, deposition and desorption occur concurrently. However, sublimation fluxes, as the desorption mechanism considered within this work, will be affected by the chemical composition of the thin film, which is in turn affected by deposition and sublimation fluxes. To model this iteratively, consecutive steps of deposition and sublimation intervals of 10 s each were considered. Thin film composition and phase formation were updated in between each step, where the individual thermochemical equilibrium calculations were performed using FactSage macros. Other interval lengths could have been used as well, since this is primarily a numerical parameter. During deposition, 3.676x10$^{-4}$ mol Mg and 7.352x10$^{-5}$ mol Ca (Mg/Ca ratio of 5) are added to the thin film composition, thus simulating the experimentally reported deposition flux using a composite Mg-Ca target exhibiting a Mg/Ca ratio of 5[12]. Using this updated composition, equilibrium phase formation and the corresponding vapor pressures of Mg and Ca, as the dominating gas species, were calculated. By applying Hertz-Knudsen equation under the assumption of perfect vacuum[25,26]

$$\dot{n}_i = \frac{\alpha_i p_i N_A}{\sqrt{2\pi M_i RT}},$$  (1)

where $\alpha_i$ is the desorption coefficient, $p_i$ the equilibrium vapor pressure, $N_A$ Avogadro constant, $M_i$ the molar mass, $R$ the ideal gas constant and $T$ the temperature, the sublimation flux $\dot{n}$ and consequently the sublimated amount of the species $i$ (Mg and Ca) within 10 s was determined individually. For these calculations, the desorption coefficients $\alpha_i$ of Mg from hcp Mg and Ca from fcc Ca were taken to be 1, respectively, while the Mg and Ca desorption coefficients from Mg$_2$Ca were assumed to be 2/3 and 1/3, respectively, corresponding to their mole fractions. Due to the low solubilities of Mg in fcc Ca and Ca in hcp Mg, these sublimation modes are not considered in the calculations. The remaining amounts of Mg and Ca after the sublimation step were subsequently used as input for the next deposition interval. Modelling of the phase formation was performed until convergence with a constant Mg/Ca ratio (fluctuations below 0.001) for 10 consecutive calculations was achieved. The final step was always a sublimation step. In analogy to the experiments[12], simulations were performed for temperatures between 20 °C and 420 °C.

### 2.2 Experimental methods

To further investigate the stoichiometry range of the Laves phase Mg$_2$Ca experimentally, combinatorial Mg-Ca thin films were synthesized in a high vacuum laboratory-scale deposition chamber by direct-current magnetron sputtering using circular elemental Mg (99.95% purity) and Ca (99.5% purity) targets with a diameter of 50 mm at base pressures below 2x10$^{-5}$ Pa. The Ca target cleaning procedure is described elsewhere[27]. Thin films were synthesized at room temperature (without intentional heating) at an Ar pressure of 0.4 Pa onto stationary Si(100) substrates resulting in the formation of a composition gradient along the substrate. The deposition time was 10 min applying constant Mg and Ca target powers of 200 W and 140 W, respectively. These films were subsequently annealed in the deposition chamber for 1 h to trigger crystallization at varying heater temperatures of up to 300 °C, thus, denoted as annealing temperature. The corresponding substrate temperature, even though not directly measured, is expected to be close to the set heater temperature as it will be discussed in the results section of this paper.

Correlative chemical and structural analyses were performed by energy dispersive X-ray spectroscopy (EDX) and X-ray diffraction (XRD) to study the composition-induced phase formation along the chemical gradient. EDX was performed using an EDAX Genesis 2000 analyzer implemented in a JEOL JSM 6480 scanning electron microscope. The acceleration voltage was set to 10 kV and the measurement time was 120 s at a magnification of 1000x. For the structural analysis of the films, a Bruker D8 General Area Detection Diffraction System (GADDS) with Cu K$\alpha$ radiation was employed. The voltage and current were set to 40 kV and 40 mA, respectively. The angle of incidence was fixed at 15° whereas the 2θ range was measured from 15° to 75°. Peak positions based on the obtained diffractograms were determined by using the TOPAS software (version 3) employing a pseudo-Voigt II function. The lattice parameters *a* and *c* of the hexagonal C14 structure of Mg$_2$Ca were calculated using the CellCalc software (version 2.10) considering the (110), (103), (112), (201), (110), (103), (201), and (004) reflections.

# 3. Results and Discussion

To critically appraise the quality of the model, the calculated chemical composition expressed as the Mg/Ca ratio compared to the reported experimental results for sputtered Mg-Ca thin films at varying substrate temperatures[12] is shown in **Figure 1** a).

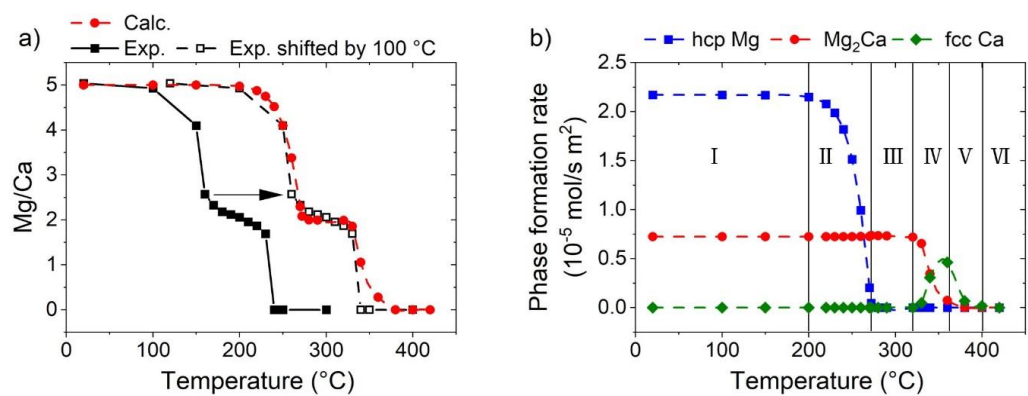

**Figure 1**: a) Calculated temperature-dependent Mg/Ca ratio based on the applied model compared to the experimental literature data for deposited Mg-Ca thin films at varying temperatures as reported[12] (full black symbols) and shifted by 100 °C (open black symbols). b) Calculated phase formation rate for hcp Mg, Mg$_2$Ca and fcc Ca as a function of the temperature separated into distinct regions with characteristic trends (I-VI).

There is a qualitative agreement in the overall trend where both the experiment and the calculation indicate a two-step loss of Mg from the thin film with a plateau spanning over a temperature range of approximately 60 °C around a Mg/Ca ratio of 2. However, a pronounced deviation in the respective temperature ranges is obtained. Experimentally, a decrease in Mg/Ca ratio is observed above 100 °C while the onset of Mg-loss is calculated to be above 200 °C. The discrepancy between calculation and experiments may be caused by various reasons: Considering that the temperature calibration is conducted with a reference sample without plasma[12], which is expected to result in an additional heat impact on the substrate, it is expected that the actual substrate temperature during synthesis is higher. Additionally, an imperfect contact between thermocouple and substrate resulting in an underestimation of the synthesis temperature cannot be excluded. Therefore, it is noteworthy that increasing the experimental substrate temperature by 100 °C results in an excellent quantitative agreement between experiment and calculation. Hence, it is reasonable to assume that the experimental substrate temperature is in fact approximately 100 °C higher and, therefore, temperatures used in the following text refer to the temperature used in the simulation or to the experimental calibration temperatures [12] shifted by +100 °C.

The composition-temperature trend, shown in Figure 1 a), can be characterized by six distinct regions which will be discussed in the following considering the predicted phase formation rate as a function of the synthesis temperature as shown in Figure 1 b).

In region I, extending up to a temperature of 200 °C, both hcp Mg and intermetallic Mg₂Ca are forming at a temperature-independent rate, resulting in a film with an overall Mg/Ca ratio of 5 (see also Figure 1 a)). Between 200 and 270 °C (region II), the hcp Mg phase formation rate is decreasing rapidly with increasing temperature while the Mg₂Ca formation rate remains constant. A further increase in temperature up until 320 °C (region III) yields the formation of phase pure Mg₂Ca at a temperature-independent rate. Region IV, ranging up to 360 °C, is characterized by a decreased phase formation rate of Mg₂Ca while elemental fcc Ca starts to form. Between 360 and 380 °C (region V), the formation of Mg-containing phases ceases but phase pure fcc Ca is still forming at a reduced formation rate. At even higher temperatures (region VI), all deposited species are sublimating and no film is forming.

This obtained temperature trends in composition and phase formation can be understood considering the equilibrium vapor pressures of Mg and Ca (exemplarily depicted for a constant temperature of 300 °C in Figure 2 c)) and the corresponding sublimation and deposition fluxes of Mg (Figure 2 a)) and Ca (Figure 2 b)) as calculated with the model.

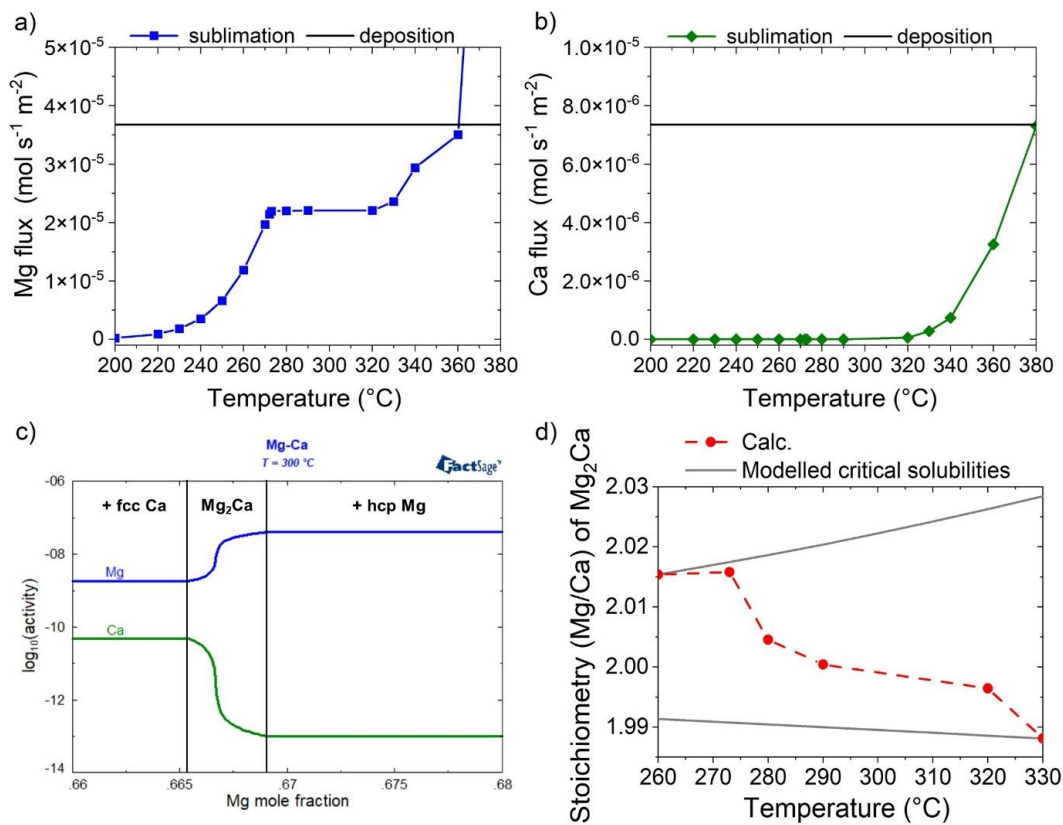

**Figure 2**: The calculated sublimation and deposition fluxes of Mg (a)) and Ca (b)) between 200 and 380 °C. c)The Mg (blue) and Ca (green) vapor pressures (=activity) at 300 °C as a function of the Mg

mole fraction and, thus, phase formation as indicated by the vertical solid black lines separating regions of $Mg_2Ca$ + fcc Ca, phase pure $Mg_2Ca$, and $Mg_2Ca$ + hcp Mg. d) The calculated stoichiometry, as Mg/Ca ratio, of $Mg_2Ca$ focused on the temperature region with predicted phase pure $Mg_2Ca$ formation (270 – 320 °C). The solid lines (grey) indicate the critical equilibrium solubilities as modelled in the FTlite database used for the conducted thermochemical equilibrium calculations.

The sublimation fluxes in Figure 2 a) and b) are affected by both temperature and composition of the film, primarily through the changes in equilibrium vapor pressures of Mg and Ca according to Eq. 1. For constant composition, the equilibrium vapor pressures generally increase with increasing temperature. At constant temperature, there is a constant vapor pressure of Mg for all compositions in the two-phase region hcp Mg + $Mg_2Ca$ and Ca + $Mg_2Ca$, where the latter one is significantly lower (see Figure 2 b)). In contrast, in the (temperature dependent) composition range for phase pure $Mg_2Ca$ formation, the vapor pressure of Mg decreases continuously and drastically when decreasing the Mg/Ca ratio (corresponding to a decreasing Mg mole fraction). Considering this, the emergence of the various temperature regions can be explained as follows: Up to a temperature of 200 °C (region I), no relevant sublimation is observed due to the low vapor pressure of Mg at these temperatures. Thus, in this temperature range the deposited ratio of Mg/Ca is conserved in the growing film. However, between 200 and 270 °C (region II), sputtered Mg is partly sublimating from the surface of the growing film while the constant incoming deposition flux is still exceeding the Mg solubility in intermetallic $Mg_2Ca$, so hcp Mg forms as a second phase. Thus, in this range the net film forming flux (deposition flux – sublimation flux) of Mg is higher than two times the net film forming flux of Ca (see Figure 2 a) and b)). However, the previously discussed pronounced temperature dependency in hcp Mg phase formation rate is driven by the exponential increase in Mg sublimation flux. The temperature regime between 270-320 °C (region III) in which phase pure $Mg_2Ca$ forms, is characterized by the fact that the net film forming flux of Mg has become approximately two times the net film forming flux of Ca, as shown in Figure 2c) and d). Hence, all Mg that cannot be incorporated in the growing $Mg_2Ca$ film is sublimating from the surface. The obtained almost constant Mg sublimation flux between 270 – 320 °C is caused by the superimposition of the temperature-induced increase and the composition-induced decrease in Mg vapor pressure (see Figure 2 c)), balancing each other out. The composition-induced decrease in Mg vapor pressure is highlighted by the calculated Mg/Ca ratio of phase pure $Mg_2Ca$ as a function of temperature (Figure 2 d)).

At 260 °C, hcp Mg still forms as a second phase and intermetallic $Mg_2Ca$ with maximum Mg solubility as modelled in the FTlite database is forming while above this temperature a continuous transition from Mg-rich $Mg_2Ca$ to Ca-rich $Mg_2Ca$ is observed with rising temperature. This transition is mainly driven, based on the thermochemical modelling in the FTlite database in which vacancies and anti-site defects on both sublattices are considered, by anti-site defects in agreement with calculated defect formation energies using density function theory[28]. Hence, the decrease in Mg concentration in phase pure $Mg_2Ca$ yields a decrease in Mg vapor pressure (Figure 2 c)) and is, thus, compensating for the temperature-induced increase in sublimation flux naturally occurring with rising temperature. With a further increase in temperature above 320 °C up to 360 °C (region IV), the vapor pressure of Mg above $Mg_2Ca$ (Figure 2 a)), and thus the Mg sublimation flux (Figure 2 a)), is rising continuously resulting in a decreased phase formation rate of $Mg_2Ca$ and in the formation of elemental fcc Ca (Figure 1 b)) since excess Ca, beyond the maximum Ca solubility in $Mg_2Ca$, is now present at the growing thin film surface. Despite the significant increase in Ca sublimation flux in this temperature regime, enhancing the decrease in $Mg_2Ca$ phase formation rate, the Mg sublimation flux is still around 10 times higher than the one from Ca (Figure 2 a) and b)) at 360 °C. Above 360 °C (region V), the Mg sublimation flux finally exceeds the deposition flux preventing the Mg incorporation into the growing film causing the formation of phase pure fcc Ca despite the continuous Mg deposition flux. The exponential increase in Ca sublimation flux, however, leads to a near-zero phase formation rate of Ca at 380 °C. At even higher temperatures (region VI), the net phase formation flux of Mg as well as of Ca approach zero inhibiting the formation of a film.

In contrast to the modelled stoichiometry range, ranging e.g. at 290 °C from 1.99 to 2.02 (see Figure 2 d)), the experimentally obtained composition range, where XRD phase pure Mg$_2$Ca is obtained, is reported to be between 1.7 to 2.2 based on the sublimation-dominated sputtering process of Mg-Ca thin films[12]. In general, controversial results regarding the stoichiometry range of Mg$_2$Ca are reported in the literature ranging from a line compound[29] to an off-stoichiometric formation[12,30,31]. To investigate the stoichiometry range of intermetallic Mg$_2$Ca, compositionally graded Mg-Ca thin films were synthesized without intentional substrate heating and were subsequently annealed in vacuum without atmosphere exposure at varying temperatures. The results of the correlative composition-structure analyses along the chemical gradient of the samples is presented in Figure 3 a).

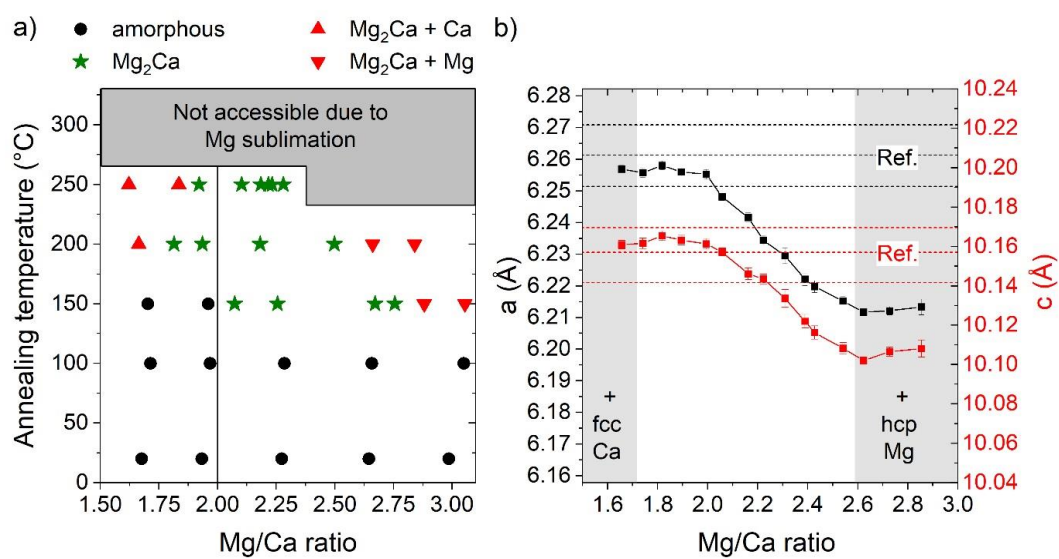

**Figure 3**: a) Phase formation map based on correlative chemical (EDX) and structural (XRD) analyses on compositionally graded Mg-Ca thin films as a function of annealing temperature and Mg/Ca ratio. (b) Determined a and c lattice parameters of hexagonal (C14) Mg$_2$Ca along the gradient of a thin film annealed at a heater temperature of 200 °C for 1 h. For comparison, lattice parameter values reported in the literature are added as dashed lines[29,32,33].

The as-deposited combinatorial Mg-Ca film and the one after annealing at around 100 °C exhibit an X-ray amorphous structure spanning over the whole Mg/Ca composition gradient ranging from around 1.6 to 3. Upon annealing at around 150 °C for 1 h, the formation of Mg$_2$Ca and Mg crystallites is obtained for a Mg/Ca ratio > 2.0, while the film with a Mg/Ca ratio < 2.0 remains amorphous potentially due to kinetically limited crystallization. A further increase in annealing temperature to 200 °C yields a crystalline film along the entire chemical gradient present on the 2'' wafer, still covering a Mg/Ca range from around 1.6 to 2.9 suggesting that sublimation does not play a major role in this temperature regime. After an annealing process at a heater temperature of 250 °C, however, the composition gradient has decreased to a Mg/Ca ratio of 1.6 to 2.3 indicating sublimation on the Mg-rich side of the deposited wafer. The here discussed temperatures indicating the transition from no substantial Mg sublimation to significant Mg loss are in good agreement with the here presented modelled onset of Mg-sublimation above 200 °C during sputtering of Mg-Ca thin films using intentional heating during deposition (Figure 1). At even higher annealing temperatures, no Mg/Ca ratio within the here presented range could be obtained attributed to severe Mg sublimation. The fully crystalline compositionally-graded Mg-Ca thin film without indications for pronounced sublimation (indicated annealing temperature of 200 °C) is consequently analyzed in finer steps by XRD to determine the $Mg_2Ca$ lattice parameters $a$ and $c$ (Figure 3 b)) along the chemical gradient.

The almost constant lattice parameters $a$ and $c$ obtained for a Mg/Ca ratio between 1.7 and 2.0 (Mg/Ca) lie within the reported lattice parameters ranges varying from 6.2514 to 6.2709 Å and from 10.1417 to 10.1696 Å, respectively[29,32,33]. For a Mg/Ca ratio of 1.7, traces of fcc Ca were obtained by XRD which vanish upon an increase in Mg concentration. It needs to be considered, however, that due to the high reactivity of fcc Ca with atmosphere (exposure required for XRD analysis), the formation of $Ca(OH)_2$ is expected[12,27]. Hence, no attempt is conducted here to determine a limit for the formation of phase pure $Mg_2Ca$ on the Ca-rich side. However, a variation in lattice parameters $a$ and $c$ for the $Mg_2Ca$ phase is considered a reliable indicator for the off-stoichiometric formation of the phase. For a Mg/Ca ratio between 2 and 2.5, a continuous decrease in lattice parameters $a$ and $c$ is obtained while a further increase in Mg concentration yields the formation of hcp Mg as a second phase as obtained based on XRD. A decrease in lattice parameters with increasing Mg concentration in $Mg_2Ca$ is in line with the atomic size ratio of 1.231 (Ca/Mg)[34] where a decrease in lattice parameters is expected upon Mg anti-site defect formation (Mg atoms on Ca site). Considering the accuracy of standardless EDX quantification, as applied here, of ±2%[35], this paper does not aim to provide a precise phase formation range of $Mg_2Ca$. Nevertheless, the results indicate an extensive (at least metastable[2,3]) phase formation range, as reported by Suzuki *et al.* who observed a solubility of around 4.5 at.% excess Mg[31] (corresponding to a Mg/Ca ratio of 2.47) which is significantly wider than modelled in the FTlite database. However, it is not ensured that the thin film is sufficiently equilibrated after annealing 1 h at 200 °C. Consequently, the obtained deviation between the calculated and modelled results in the temperature region of phase pure $Mg_2Ca$ formation are at least partly attributed to the limited stoichiometry range in the model which does not account for the extended potentially metastable solubility obtained experimentally.

## 5. Conclusions

By employing CALPHAD-based thermochemical equilibrium calculations and Hertz-Knudsen equation to describe sublimation, composition and phase formation during sputtering of Mg-Ca thin films has been modelled as function of synthesis temperature and Mg/Ca ratio. The obtained agreement between model calculations and experimental results reported in the literature suggests that the temperature-induced sublimation from the surface of the growing film is the dominating mechanism determining the composition-structure relationship during sputtering in this system. Thus, it has been demonstrated that the film composition and structure can be predicted for sublimation-dominated deposition scenarios by applying equilibrium thermochemical calculations. By explicitly considering the synthesis temperature and deposition fluxes of the individual species in the model, it is expected that the model can serve for the prediction of suitable synthesis parameters for sputtering of material systems which are prone to sublimation. This study corroborates previous studies concerning the applicability of (para-)equilibrium calculations in PVD[2–6] and is, thus, a significant step towards understanding how synthesis parameters control composition and thereby phase formation in PVD.

**Author Contributions:** Conceptualization, P.K., M.t.B., and J.M.S.; methodology, P.K. and M.t.B.; software, P.K. and M.t.B.; validation, P.K., M.t.B., S.A. and J.M.S.; formal analysis, P.K.; investigation, P.K., M.t.B., and J.M.S.; resources, M.t.B. and J.M.S.; data curation, P.K.; writing—original draft preparation, P.K.; writing—review and editing, P.K., M.t.B., S.A. and J.M.S.; visualization, P.K.; supervision, M.t.B., J.M.S.; project administration, J.M.S.; funding acquisition, J.M.S. All authors have read and agreed to the published version of the manuscript.

**Funding:** This work was supported by Deutsche Forschungsgemeinschaft (DFG) within the Collaborative Research Center SFB 1394 "Structural and Chemical Atomic Complexity—From Defect Phase Diagrams to Materials Properties" (Project ID 409476157).

**Data Availability Statement:** Data available on request from the corresponding author.

**Conflicts of Interest:** The authors declare no conflict of interest.